\documentclass[fleqn,10pt]{wlscirep}
\newcommand{\blue}[1]{{\color{black} #1}}

\newcommand{\edit}[1]{{\color{black} #1}}
\newcommand{\editt}[1]{{\color{black} #1}}

\usepackage[utf8]{inputenc}
\usepackage[T1]{fontenc}
\usepackage{placeins}
\usepackage{float}
\usepackage{xcolor}
\usepackage{amsmath}
\title{Broadband Achromatic Metalens for the Short-Wave Infrared}

\author[1,*]{Yan He}
\author[1,$\dag$]{Adetunmise C. Dada}
\affil[1]{School of Physics and Astronomy, University of Glasgow, Glasgow G12 8QQ, UK}

\affil[*]{y.he.5@research.gla.ac.uk}
\affil[$\dag$]{adetunmise.dada@glasgow.ac.uk}

\begin{abstract}
The 1.8-2.3~$\mu$m band lies within the short-wavelength infrared (SWIR) region and serves as a key operational window for a wide range of applications, including quantum sensing, molecular spectroscopy, and free-space quantum and classical optical communication. Despite its significance, optical devices operating in this band still face two major challenges: chromatic aberration across the wide spectral range and the difficulty of integration due to bulky optical elements. Metalenses are composed of subwavelength nanostructures that locally control the phase and group delay of light, enabling precise wavefront shaping and broadband dispersion compensation. These capabilities make them highly promising for use in infrared optical systems, particularly in applications such as focusing and imaging for compact integrated devices.

In this study, we propose a metalens design based on a CaF$_2$ substrate, where each nanocell consists of a single-bar silicon structure. These nanocells are periodically arranged with a 900~nm period, enabling precise control of dispersion and phase. By systematically finetuning the bar length and width, the design enables simultaneous dispersion compensation and phase modulation, achieving stable focusing performance over a broad spectral range. Finite-Difference Time-Domain (FDTD) simulations demonstrate \edit{effective suppression of chromatic aberration across 1800--2300~nm, with simulated focal-length variation within 6\% of the target value. We further analyze the polarization distribution across the focal spot and find a weak wavelength dependence of the degree of polarization (DoP), which we attribute to the spatially varying polarization state in the high-NA focal region together with the wavelength-dependent anisotropic response of the nanostructures. Meanwhile,} this design offers a compact, broadband, and high-performance approach for beam collimation and wavefront shaping in the SWIR band, showing promising potential for applications in quantum communication and sensing systems.
\end{abstract}
\begin{document}

\flushbottom
\maketitle
%
%

\section{Introduction}
The short-wave infrared (SWIR) band typically refers to the spectral range of 1--2.5~$\mu$m~\cite{Mars2010}. Sub-bands around 2.1--2.2~$\mu$m coincide with an atmospheric transmission window~\cite{Campargue2017}, where absorption by water vapor and carbon dioxide is relatively weak compared with adjacent bands; thus, SWIR light avoids strong H$_2$O/CO$_2$ absorption features and exhibits relatively low atmospheric loss and more robust atmospheric performance~\cite{Knaeps2015}. Compared with visible light, SWIR light propagates more reliably through haze and smoke~\cite{Pugliese2023}, reducing scattering and losses, and it maintains good imaging and detection capability under low light or partial obstruction~\cite{Wen2018}. In addition, some materials exhibit distinctive absorption or emission responses in this wavelength range~\cite{Attiaoui2020,Sarkar2020,Leemans2022,Wilson2015}, enhancing sensitivity and imaging quality and enabling applications in industrial inspection~\cite{Gutirrez-Gutirrez2025}, material identification~\cite{Mehrubeoglu2020}, and biomedical imaging~\cite{Yao2025}. Owing to partial penetration through some otherwise opaque materials, SWIR is also widely used in remote sensing~\cite{Abuzar2024}, meteorological observation~\cite{Ustin2024}, and atmospheric composition analysis~\cite{Jacob2022}. In optical communication, SWIR supports free-space communication and short-distance data transmission due to its robust environmental tolerance and favorable transmission characteristics~\cite{Gach2020}; in quantum information, it is promising for medium- to long-distance quantum communication, particularly for free-space links operating under complex or interference-prone conditions~\cite{Prabhakar2020}.

Despite these advantages, key SWIR optical component systems remain underdeveloped, especially at the micro-/nano-optics level. Quantum communication systems impose increasingly stringent requirements on optical components~\cite{Pirandola2020}: devices must provide stable wavefronts to suppress excess phase noise and low dispersion with minimal wavefront distortion to ensure high-fidelity transmission of quantum states. However, current SWIR-compatible compact components often fall short in wavefront control, dispersion management, and system integration adaptability, limited by large dimensions, insufficient tuning precision, and miniaturization--robustness trade-offs. Compared with visible and NIR technologies, SWIR micro-/nano devices still face challenges in design flexibility and engineering feasibility~\cite{Bianconi2020}, creating bottlenecks for integration into high-precision quantum communication systems.

\begin{figure}[t]
        \centering
        \includegraphics[width=0.9\textwidth]{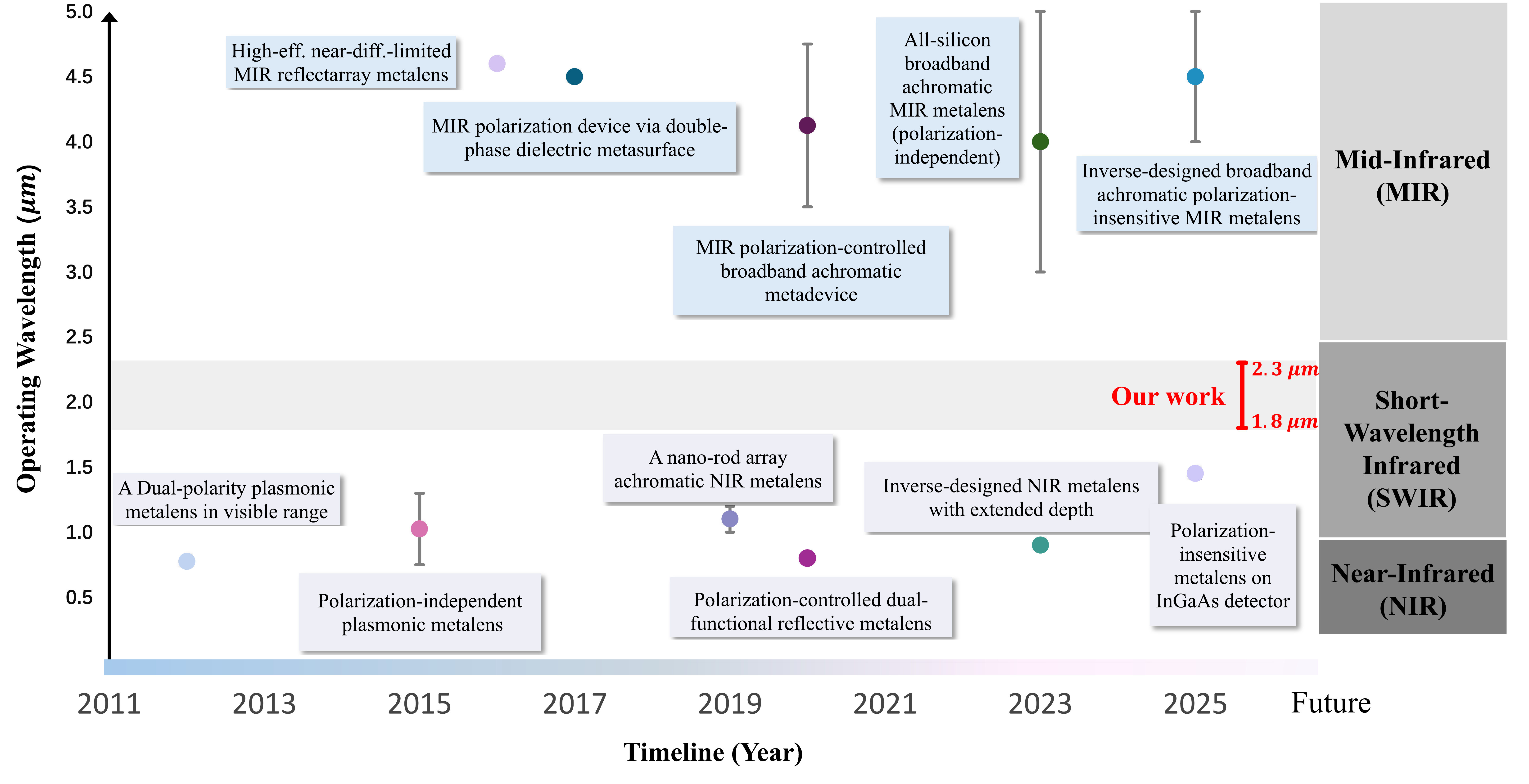}
        \caption{Timeline of representative metalens developments across the near-infrared (NIR), short-wavelength infrared (SWIR), and mid-infrared (MIR) regions. Each point marks a key study identified by its publication year and operating wavelength. The error bars indicate the range of operating wavelengths, respectively. The diagram compiles results from multiple works~\cite{Chen2012,Wang2015,Yu2019,Zhang2020,Xiao2023,Zou2025,Zhang2016,Guo2017,Ou2020,Yue2023,Xu2025}. Notably, fewer works have been reported in the SWIR band of 1.8-2.3~$\mu$m, which is the focus of this study.}
        \label{metalens_timeline}
        \end{figure}

Metalenses offer a promising route to address these limitations through compact form factors, high design flexibility, and strong capabilities in dispersion engineering and wavefront control~\cite{Khorasaninejad2017,Arbabi2015}. As metasurface-based devices, metalenses have attracted broad attention across spectral bands due to their ultrathin profiles and multifunctionality (e.g., wavefront control and dispersion compensation), enabling achromatic imaging, super-resolution imaging, polarization manipulation, and multi-wavelength optical systems~\cite{Wang2021,Chen2018,Arbabi2015}. Initially developed in the visible regime, their applications have expanded into the NIR and MIR domains (Fig.~\ref{metalens_timeline}); material platforms such as TiO$_2$~\cite{Liang2018}, Si~\cite{Huang2021}, and GaN~\cite{Chen2017}, together with diverse nanostructure designs, have enabled high transmission efficiency and functional integration across wavelengths, motivating broadband dispersion-compensating metalenses for SWIR systems and deeper integration of SWIR optics into quantum communication and photonic platforms.

These successive breakthroughs have not only facilitated the widespread integration of metalenses into conventional optical systems, but also gradually extended their application into the SWIR regime. Specifically, a number of devices operating around the telecommunication window~\cite{Zou2025,Liu2024} (approximately 1.3-1.6~$\mu$m) have demonstrated functionalities such as focusing and polarization control. Meanwhile, increasing attention has been directed toward performance metrics including dispersion engineering and system-level integration~\cite{Zhang2025-1,Zhang2025-2}.  
However, existing studies have primarily focused on narrowband or single-function optimization, leaving a gap in the development of SWIR optical devices that are \blue{both} broadband and dispersion-engineered, as well as their integration \blue{solutions}. This gap is particularly evident in the 1.8-2.3~$\mu$m~range, a critical window bridging the near-infrared and mid-infrared regions. Despite its significant potential in quantum information transmission, nonlinear optics, and \blue{bridging near-IR to mid-IR applications}, there remains a lack of systematic metalens designs and experimental validations in this band. \blue{To address this gap, we propose a broadband achromatic metalens operating in the 1.8–2.3~$\mu$m SWIR band.} Compared to previous demonstrations such as \cite{Zou2025} which employed multilayer intermediate structures including SiO$_2$ transition layers to achieve achromatic focusing, our design adopts a unified phase compensation strategy based on geometric and Pancharatnam-Berry (PB)-phase tuning. This simple structure 
 facilitates integration with standard processes, thereby enhancing the scalability and practicality of the device across the SWIR band.

We numerically validate an achromatic metalens with stable focusing across 1.8--2.3~$\mu$m and quantify focal‑length stability ($\leq 6\%$) and efficiency under the reported design parameters, achieved by jointly engineering phase and dispersion through geometric and PB phase tuning.
Most related works concentrate on a single, narrow-wavelength band, or limited-bandwidth performance, without offering integrated solutions that simultaneously provide broadband coverage and dispersion compensation. This gap motivates deeper investigation in this spectral range. Here, we target the underexplored 1.8--2.3~$\mu$m SWIR window and present a single-layer metalens platform on CaF$_2$ using a single-bar Si nanostructure library. By co-optimizing geometric parameters to simultaneously control transmission phase and dispersion, we achieve broadband achromatic focusing in full-wave simulations. We further analyze focusing stability, efficiency, and polarization behavior across the band, and discuss fabrication feasibility within current lithographic limits.

\subsection{Metalens Principle and Design Framework}
\begin{figure}[h]
\centering
\includegraphics[width=0.9\textwidth]{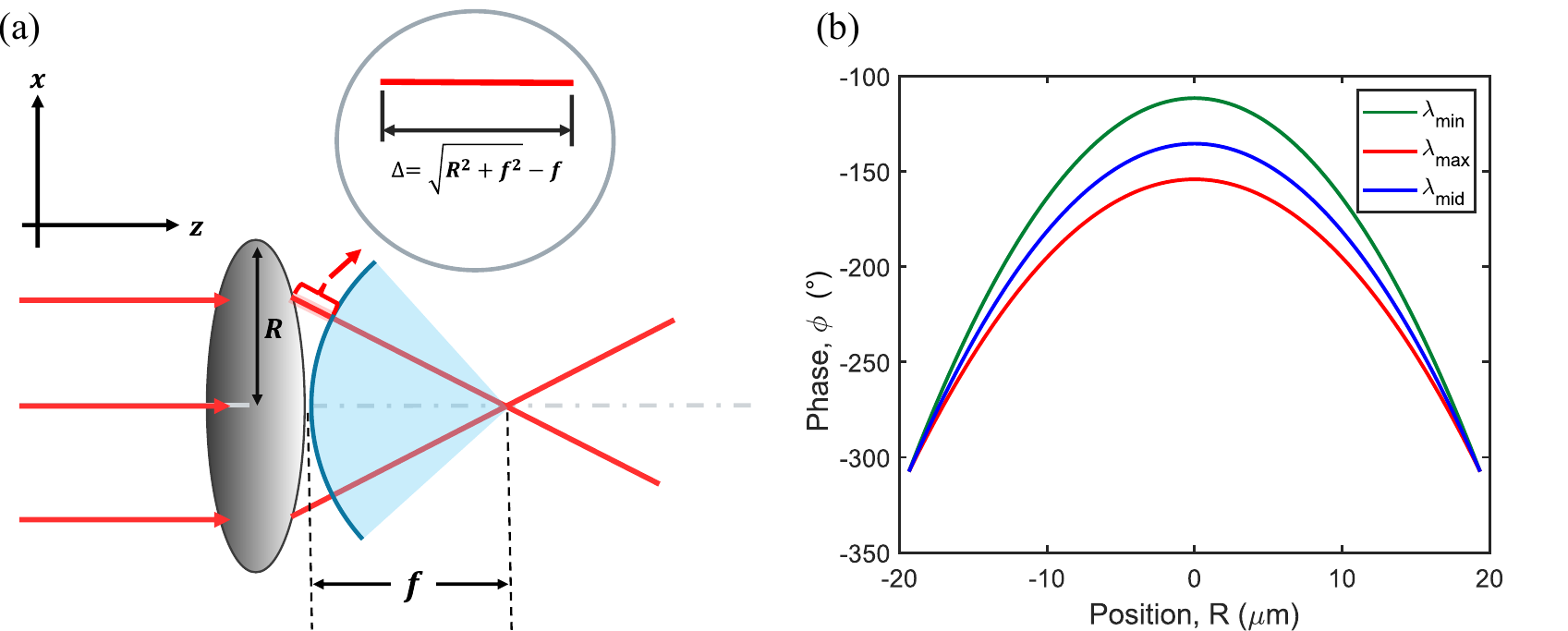}
        \caption{(a) Schematic diagram showing the ideal wavefront distribution for a metalens. To focus parallel incident light to a single focal point at distance $f$, the metasurface must introduce a spatially varying phase delay to compensate for the optical path difference between the center and off-axis positions. The required phase compensation at each radial distance $R$ from the optical axis is given by $\Delta = \sqrt{R^2 + f^2} - f$, ensuring that all rays interfere constructively at the focus~\cite{Chen2018}. (b) Calculated phase distribution of the metalens along the radial direction for three different design wavelengths ($\lambda_{\text{min}} = 1.8\,\mu\text{m}$, $\lambda_{\text{mid}} = 2.05\,\mu\text{m}$, and $\lambda_{\text{max}} = 2.3\,\mu\text{m}$). The curves indicate the required phase compensation to achieve focusing at the designed focal length, showing the variation of phase with respect to wavelength. \editt{Note that the plotted phase profiles include the additive offset $C(\lambda)$, which sets the phase reference (e.g., $\phi(R=0,\lambda)$).}}
\label{fig:Ideal wavefront}
\end{figure}

The focusing functionality of conventional lenses primarily relies on the geometrical thickness distribution across different radial positions, which introduces varying optical path lengths and consequently provides the necessary propagation phase compensation for the incident light. Through proper design, light waves from different positions accumulate appropriate phase delays, enabling constructive interference at the focal point and thus forming a clear focusing effect. In contrast, metalenses employ subwavelength-scale nano units to precisely manipulate the spatial phase distribution of the transmitted light, thereby actively shaping the wavefront. To achieve ideal focusing performance, the spatial phase at each position on the metalens must satisfy the following distribution, as illustrated in Fig. \ref{fig:Ideal wavefront}(a)~\cite{Chen2018}:
\begin{equation}
        \varphi_{\mathrm{ideal}}(R, \lambda) = -\frac{2\pi}{\lambda} \left(\sqrt{R^2 + f^2} - f \right)~\edit{+~C(\lambda)}, 
\label{eq:1}
\end{equation} 
 \edit{where $R$ represents the radial position (with its unit determined by the specific design requirements). $f$ is the designed focal length, $\lambda$ is the operating wavelength, and $C(\lambda)$ is an additive, position-independent phase constant that sets the phase reference at each wavelength (e.g., chosen such that $\varphi_{\mathrm{ideal}}(0,\lambda)=0$).}

\editt{While the first term ensures the required spatial phase gradient for focusing, the additive phase term $C(\lambda)$ is independent of position and therefore shifts the phase profile by a constant offset (i.e., it sets the phase reference at each wavelength, e.g., $\phi(R{=}0,\lambda)$) without affecting the focusing condition, which depends on the spatial phase gradient. However, $C(\lambda)$ plays an important role in controlling the overall phase--frequency relationship and hence the group delay of the transmitted wavefront.}
In practical metalens design, structural dispersion and local geometric constraints often lead to variation in the output phase with varying incident wavelengths. 
Therefore, in addition to satisfying the phase requirements at a single wavelength, it is also essential to comprehensively consider the phase variation trends across different wavelengths, as illustrated in Fig. \ref{fig:Ideal wavefront}(b).

This work applies a compensation design method, which involves introducing a wavelength-dependent compensation phase term\cite{Aieta2015}. 
The phase compensation relationship at different wavelengths can be expressed as~\cite{Wang2018,Guo2022} \edit{\\$\varphi_{\text{ideal}}(R, \lambda) = \varphi(R, \lambda_{\text{max}}) + \Delta\varphi(R, \lambda)$, with a compensation phase $\Delta \varphi(R, \lambda) = -\left[ 2\pi \left( \sqrt{R^2 + f^2} - f \right) \right] \left( \frac{1}{\lambda} - \frac{1}{\lambda_{\text{max}}} \right)+ C(\lambda)$ .}
\begin{figure}[t]
  \centering
  \includegraphics[width=0.6\textwidth]{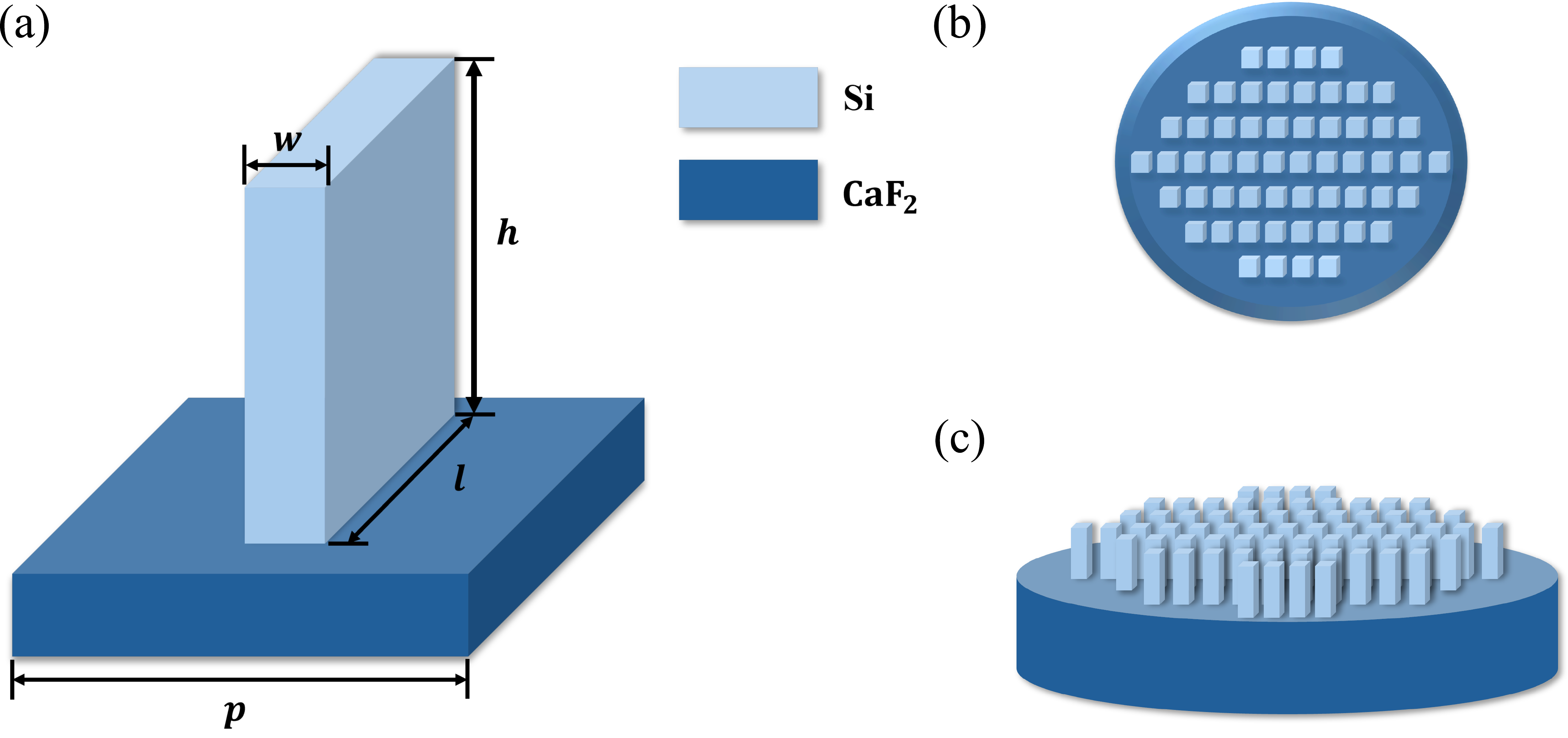}
  \caption{Unitcell profile and structure of Metalens. (a) Schematic of the unit cell geometry used in the metalens design. The rectangular dielectric nanopillar has a length \textit{l}, width \textit{w}, and height \textit{h}, and is placed on a dielectric substrate. (b) Top view of the metalens layout, showing the spatial arrangement of rectangular meta-atoms with varying geometries across the aperture. (c) Perspective view of the complete metalens composed of anisotropic nanopillars, where each unit is tailored \editt{by varying geometries ($l$, $w$) and rotation angles ${\alpha}$ across the aperture (with $\alpha$ defined as the in-plane rotation angle of the rectangular nanopillar about its vertical axis)} to locally modulate the transmitted wavefront.}
  \label{unitcell-3Dlayout}
  \end{figure}
\begin{figure}[h]
  \centering
  \includegraphics[width=1\textwidth]{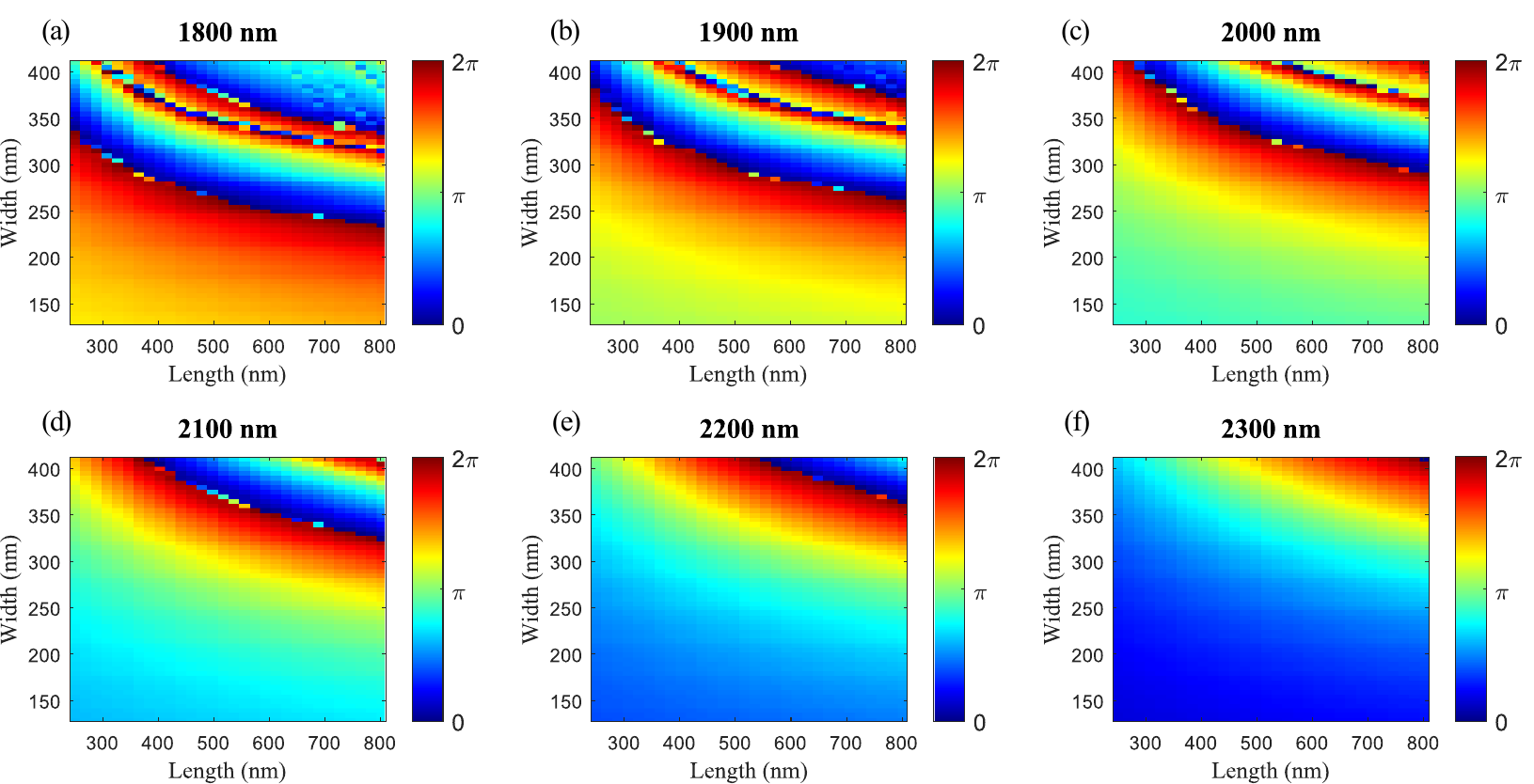}
  \caption{Phase response maps of the unit cell under different wavelengths (1800~nm~to 2300~nm). The phase distribution is plotted as a function of nanostructure length and width, showing a smooth phase variation across the SWIR band and achieving full 0-2$\pi$ modulation. \editt{The plotted phase corresponds to $\arg\!\left(t_{L}-t_{S}\right)$, where $t_{L}$ and $t_{S}$ denote the complex transmission coefficients along the long and short axes, respectively}.}
  \label{phase distribution}
  \end{figure}

In the design of metalenses, the output phase imparted by each unit structure can typically be expressed as~\cite{Wang2018,Chen2019}:
\begin{equation}
    \varphi_{\text{lens}}(R,\lambda)
    = \varphi_{\mathrm{ideal}}(R,\lambda)
    + \Delta \varphi_{\text{struct}}(R,\lambda),
\label{eq:4}
\end{equation}
where $\Delta \varphi_{\text{struct}}(R,\lambda)$ represents the additional phase introduced by the specific meta-atom (including propagation and PB contributions). 
 It is worth noting that the wavelength-dependent variation of this phase distribution essentially reflects the group delay (GD) characteristics of the system. Group delay describes the derivative of phase with respect to frequency (or wavelength), which can be expressed as:
        \begin{equation}
        \tau_{g}(R) = \frac{d\varphi(R,\lambda)}{d\omega} = - \frac{d\varphi(R,\lambda)}{d\lambda} \cdot \frac{d\lambda}{d\omega},
\label{eq:6}
        \end{equation}
where $\tau_{g}(R)$ describes the time delay for different frequency components at position $R$. The output phase $\varphi(R,\lambda)$ depends on geometric path difference, structural dispersion, and PB phase. $\omega = 2\pi c/\lambda$ is the angular frequency, and the derivatives characterize phase and dispersion responses of the system.

\vspace{1em}Therefore, broadband metalens design requires not only static phase compensation at individual wavelengths, but also comprehensive consideration of the group delay distribution to ensure temporal consistency of the wavefront. By optimizing the structural parameters and compensation phase distribution, the group delay curve can be effectively flattened, thereby reducing focus shifts and wavefront distortions induced by dispersion, and ultimately achieving true broadband, high-quality imaging performance.~\blue{The overall design approach enables broadband dispersion management and achromatic focusing, supporting consistent optical performance across the spectrum.} However, the local tuning capability of geometric dimensions is inherently limited. As a result, relying solely on propagation phase compensation cannot fully satisfy the comprehensive design requirements of broadband metalenses, particularly in terms of wavefront consistency and chromatic aberration correction. To further enhance the overall phase control capability and dispersion suppression performance of the metalens under broadband operation, this work introduces the PB phase as an \blue{auxiliary phase modulation approach in the design}.
        
The PB phase originates from the change in the polarization state of light, and essentially manifests as an additional phase introduced through the rotation of the nanostructure. Specifically, when incident circularly polarized light passes through an anisotropic unit structure, the corresponding transmission matrix can be expressed as:
        \begin{equation}
        t = \frac{t_L + t_S}{2} 
        \begin{pmatrix}
        1 \\i
        \end{pmatrix}
        + \frac{t_L - t_S}{2} e^{i2\alpha}
        \begin{pmatrix}
        1 \\-i
        \end{pmatrix},
\label{eq:7}
        \end{equation}
where $t_L$ and $t_S$ represent the transmission coefficients of the structure along the long and short axes, respectively, and $\alpha$ denotes the rotation angle of the structure. \edit{For Eq. \eqref{eq:7} the incident field is assumed to be left-handed circularly polarized and can be represented by the Jones vector $(1,i)$.} By adjusting the rotation angle $\alpha$ of the structure, controllable polarization-dependent phase compensation can be flexibly introduced without altering the overall dimensions or material parameters of the structure. The PB phase exhibits excellent broadband response and low dispersion characteristics, which effectively overcome the design limitations imposed by structural dispersion. In combination with propagation phase modulation, it significantly enhances the overall chromatic aberration compensation capability and wavefront control performance under broadband operation.

\section{Design and Results}

A single-bar nanocell structure based on silicon (Si) nanopillars on a calcium fluoride (CaF$_2$) substrate was designed to achieve efficient transmission modulation and precise phase control in the SWIR region. Calcium fluoride exhibits a low refractive index and excellent infrared transparency, while silicon offers a high refractive index and low absorption loss within this wavelength range. \blue{This material combination ensures a broad phase modulation range and stable focusing behavior over the operating bandwidth, while maintaining a reasonable transmission efficiency.}

\blue{On the basis of the established material configuration, we systematically designed the nanocell geometry for left circularly polarized (LCP) input light.} The unit cell period was set to $p=900$~nm, and the pillar height was chosen as $h=2~\mu$m~to minimize near-field coupling between adjacent elements and maintain \blue{wavefront shaping capability} within the operating wavelength range. \blue{In addition,} the nanocell length $l$ was varied from 250 to 800~nm, while the width $w$ was tuned from 130 to 410~nm. \blue{Using the above approach, we carried out a multi-wavelength phase optimization. Specifically, we constructed a phase library via full-wave simulations for our nanocell across the geometry parameter space~(\textit{l}, \textit{w}, $\theta$) at the design wavelengths 1.8-2.3~$\mu$m. We then selected appropriate parameters for each radial position such that the phase condition Eq.~\eqref{eq:1} \editt{[with compensation]} is approximately satisfied for all wavelengths. This was followed by finetuning~(e.g., $\theta$) to improve broadband phase consistency.} Geometric parameter tuning is combined with rotation-based phase modulation to achieve precise control of the transmitted wavefront. This approach enables the metalens to meet the design requirements for focusing and wavefront reconstruction. Fig. \ref{unitcell-3Dlayout} illustrates the schematic of the proposed nanocell, its spatial arrangement within the metalens, and the overall three-dimensional structure of the device.

Subsequently, we further analyzed the transmission phase response of the designed nanocell in the target SWIR band to evaluate whether a complete 2$\pi$ phase modulation can be achieved. \edit{Here, the transmission phase corresponds to the phase difference between the responses along the long and short axes of the nanostructure, i.e., the phase of $t_L - t_S$.} Fig.~\ref{phase distribution} shows the simulated transmission phase distribution as a function of the nanocell length \textit{l} and width \textit{w} at different wavelengths (1800-2300~nm). The colors in the maps represent the phase delay provided by nanocells with different geometric dimensions. Within the designed parameter range (\textit{l}=250-800~nm, \textit{w}=130-410~nm), the nanocell provides a continuous phase coverage from 0 to 2$\pi$. The phase response remains smooth and continuous as the geometric dimensions are varied. Therefore, this structure can serve as a fundamental building block for broadband metalens devices, providing the foundation for the subsequent analysis and optimization of phase response and \blue{focus stability over a large wavelength range}.

\begin{figure}[H]
  \centering
  \includegraphics[width=0.9\textwidth]{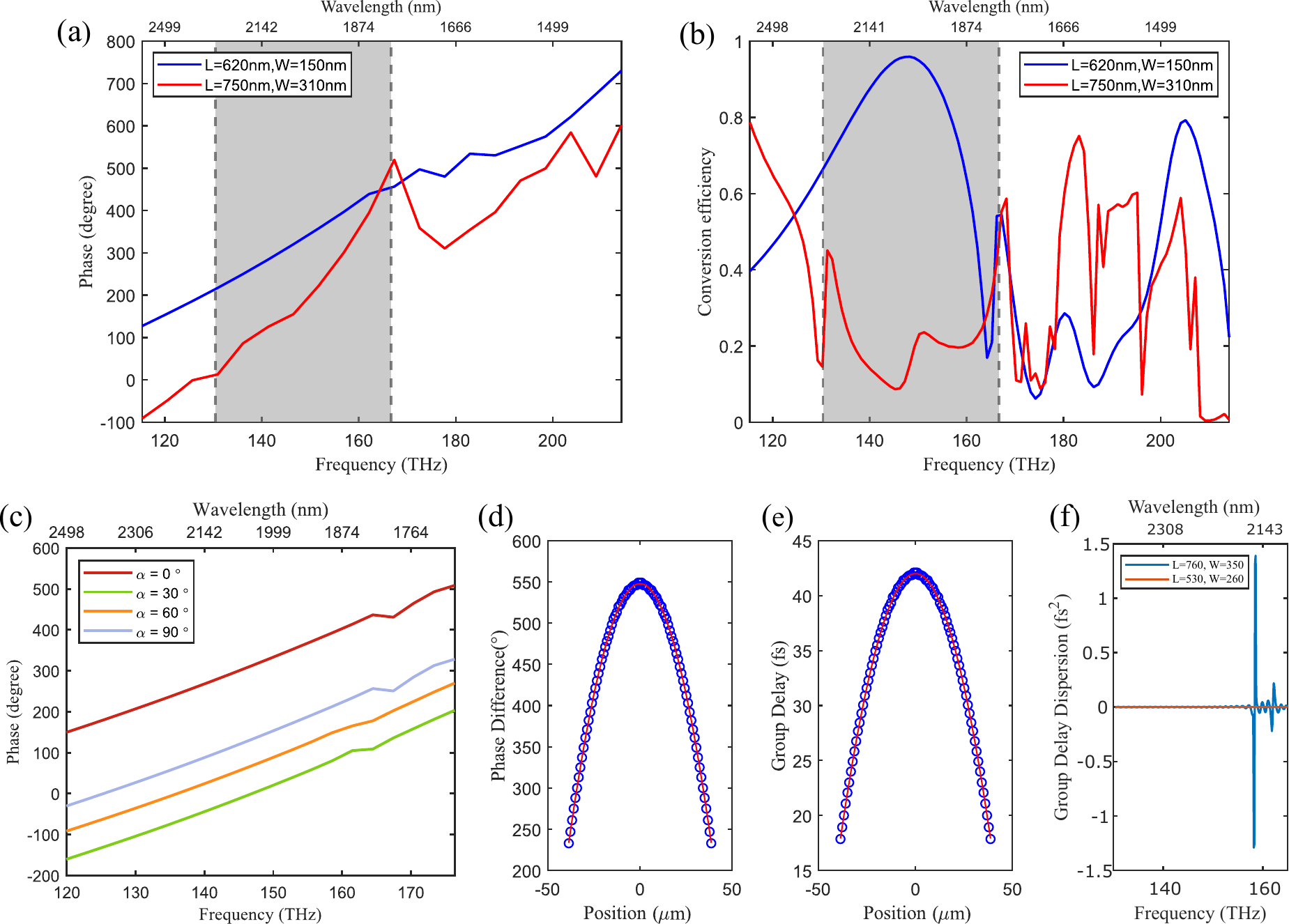}
\caption{Simulated phase response (a) and conversion efficiency (b) of rectangular dielectric nanopillars as functions of frequency for two sets of geometrical parameters (L=620~nm, W=150~nm~and L=750~nm, W=310~nm). \edit{Panels (a)–(b) show representative unit-cell responses from the parameter sweep (illustrative examples) and are not necessarily selected into the final simulated metalens design.} The shaded region highlights the target spectral range of 133-160~THz (1800-2300~nm). (c) Simulated transmission phase of the designed rectangular nanocell under different polarization angles ($\alpha = 0^{\circ}$, $30^{\circ}$, $60^{\circ}$, $90^{\circ}$). The phase shows broadband frequency-dependent variation and nearly parallel trends across polarizations, indicating stable group delay and low polarization sensitivity. Slight slope changes near 160-170 THz ($\approx$1.76-1.88~$\mu$m) have negligible impact on phase design and metalens optimization. \edit{(d) Required phase difference between the longest and shortest design wavelengths and (e) the corresponding group delay across the radius of metalens. The red lines denote the target values, and the blue markers denote the values realized by the selected nanocells. (f) Group-delay dispersion (GDD) as a function of frequency (133-160 THz) for two representative nanocells which are located near the center (L=760nm, W=350nm) and near the edge (L=530nm, W=260nm) of the designed metalens. The GDD is obtained from the unwrapped spectral phase, $GDD = \frac{d^2\phi}{d\omega^2}$.} }
\label{phaseconvsim3d}
  \end{figure}
  
  \begin{figure}[h]
  \centering
  \includegraphics[width=0.8\textwidth]{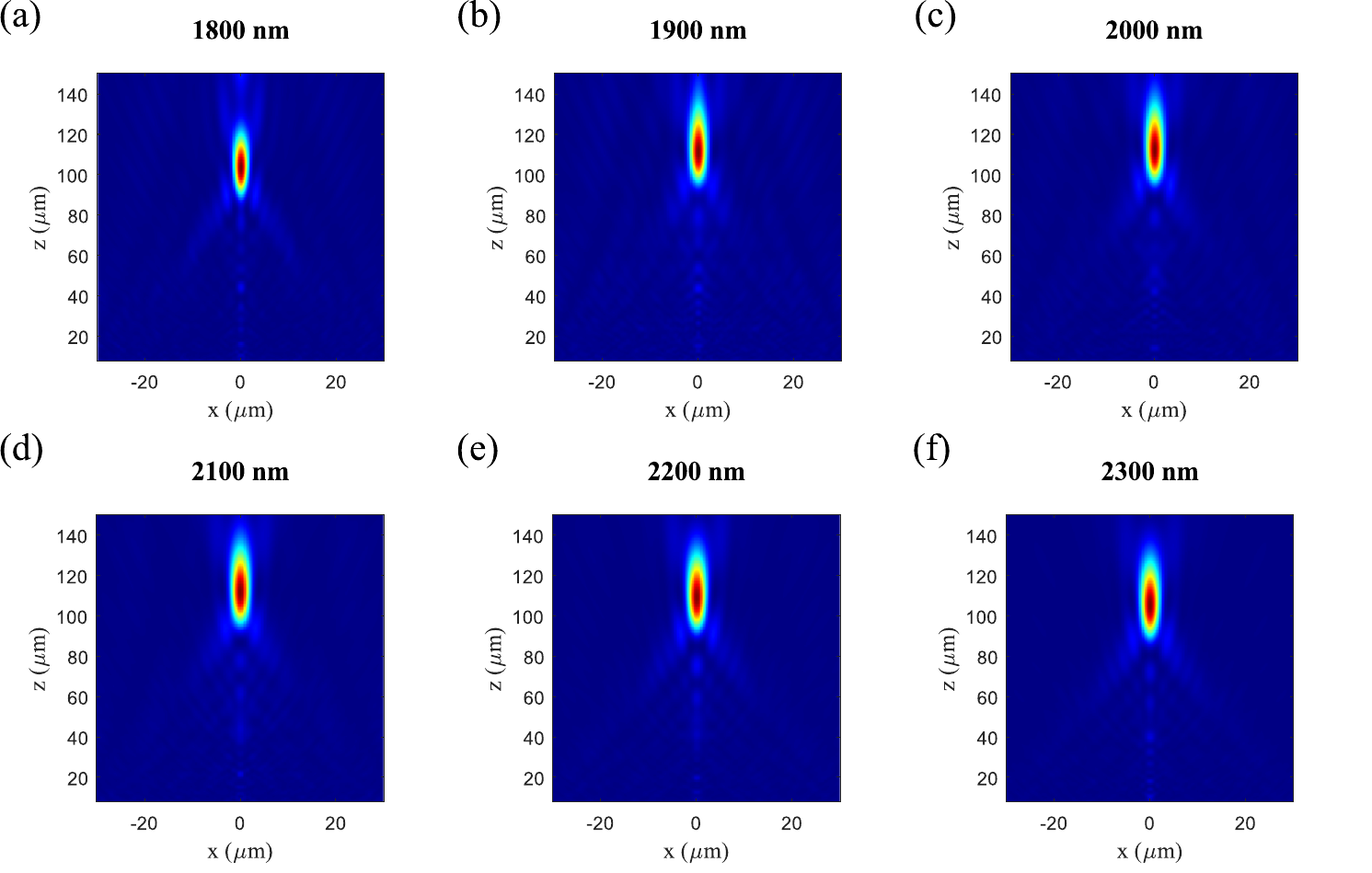}
  \caption{Simulated electric field intensity distributions ($|E|^2$) in the X–Z plane at different wavelengths (1800–2300~nm) under normal incidence, indicating stable focal positioning across the spectral range.
  }
  \label{farfield_image_allband}
  \end{figure}

 \begin{figure}[h]
  \centering
  \includegraphics[width=\textwidth]
  {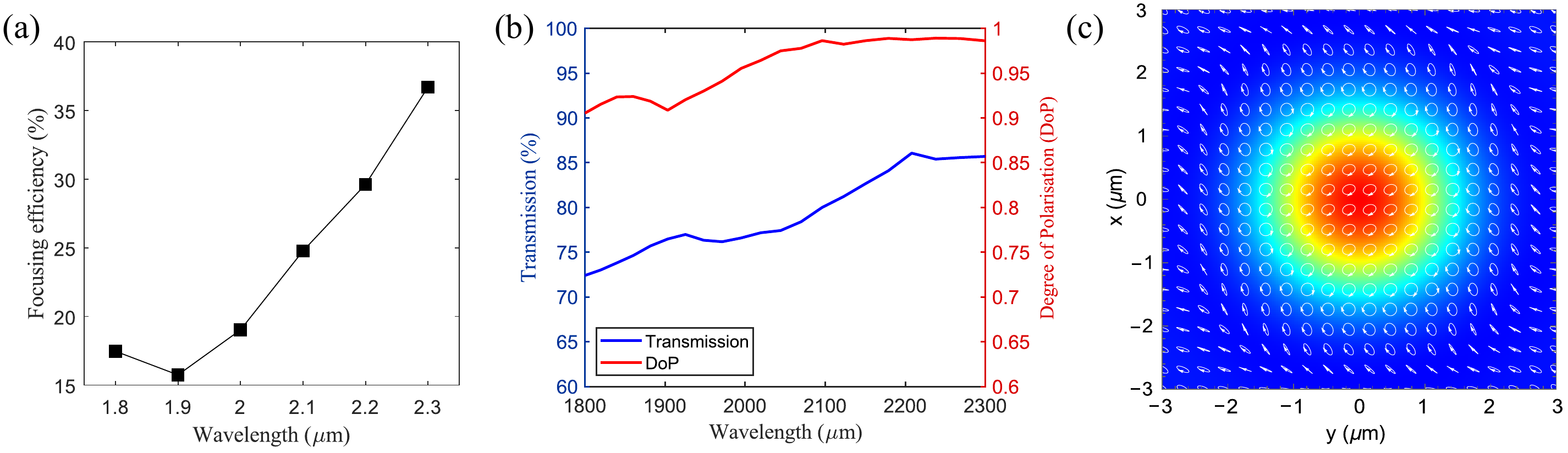}
  \caption{(a) Simulated focusing efficiency of the metalens as a function of wavelength from 1.8~$\mu$m to 2.3~$\mu$m. The efficiency increases gradually with wavelength, reaching approximately 38\% at 2.3~$\mu$m. The trend indicates improved focusing performance at longer wavelengths within the SWIR band. (b) Simulated transmission (blue curve, left axis) and DoP (red curve, right axis) of the all-band metalens across the 1800-2300~nm wavelength range. The results reveal a trade-off between achieving high transmission and preserving polarization purity across the spectrum. \editt{(c) On-axis point-spread function (PSF) at $\lambda = 1800~\mathrm{nm}$, shown as the simulated focal-plane intensity distribution at the design focus under plane-wave illumination; the overlaid vectors show the corresponding polarization distribution across the focal region.}}
  \label{focusing_efficiency_allband}
  \end{figure}

To further evaluate the optical performance of the designed nanocells in the SWIR band, we performed frequency-dependent simulations of their transmission phase and the conversion efficiency from left-handed circularly polarized (LCP) light to right-handed circularly polarized (RCP) light. Figure~\ref{phaseconvsim3d}(a)--(b) shows the simulated phase delay and LCP$\rightarrow$RCP conversion efficiency for two representative geometries (L=620~nm, W=150~nm and L=750~nm, W=310~nm) as a function of frequency. The shaded region indicates the target range of 133--160~THz (1.8--2.3~$\mu$m).  As shown in Fig.~\ref{phaseconvsim3d}(c), the transmission phase was simulated for incident polarization angles $\alpha = 0^{\circ}$, $30^{\circ}$, $60^{\circ}$, and $90^{\circ}$. The phase varies with frequency for different polarization states, confirming PB-phase-based modulation. Although the absolute phase differs among angles, the phase--frequency curves remain nearly parallel, indicating a stable group-delay trend. Minor slope variations appear around 160--170~THz ($\approx$1.76--1.88~$\mu$m), implying weak polarization-dependent dispersion; however, these effects are negligible and do not in practice limit  PB-phase-based phase distribution design or metalens optimization. \edit{As summarized in Fig.~5(d)--(e), we report the target and realized phase difference between the longest and shortest design wavelengths, together with the corresponding group-delay profile across the metalens radius (red: target; blue: selected nanocells).} \edit{Within the target band, any curvature in the phase--frequency response corresponds to nonzero (residual) group-delay dispersion (GDD), quantified as $GDD = \frac{d^2\phi}{d\omega^2}$. To quantify this effect for the implemented design, we extract and report  in Fig.~\ref{phaseconvsim3d}(f) the residual GDD over 133--160~THz for two representative nanocells selected from those used in the simulated metalens. 
The residual (nonzero) GDD is consistent with the remaining chromatic variation observed in the full-metalens simulations (i.e., the small focal-length shift across 1.8--2.3~$\mu$m).} We then analyzed the PB phase behavior.

Based on the material platform and optimized nanocell library, we built a full 3D metalens model in Lumerical FDTD to evaluate performance under realistic electromagnetic conditions. The metalens has an effective aperture radius of $R=38.7~\mu$m and focal length $f=100~\mu$m (NA=0.39). Nanopillars are arranged on a square lattice with period $p=900$~nm; each site selects a geometry $(l,w,\theta)$ from the pre-established library to implement the target focusing phase profile $\phi(r)$. 

To assess broadband focusing, we simulated the normalized electric-field intensity ($|E|^2$) distributions in the X--Z plane under normal incidence for wavelengths 1800--2300~nm (Fig.~\ref{farfield_image_allband}). Across the band, the focal spots remain well confined and close to the designed focal length of 100~$\mu$m, with longitudinal shifts within 6\% ($\pm$6~$\mu$m). Specifically, the focal length varies from 100~$\mu$m at $\lambda=1800$~nm to 105.8~$\mu$m at $\lambda=2300$~nm. The focusing profiles also remain laterally stable, indicating effective phase matching and chromatic aberration correction: no significant lateral chromatic aberration is observed, the focal spot stays centered on the optical axis, and the focal-spot size (FWHM) shows only minimal variation from 1800~nm to 2300~nm.   
 Based on the preceding analysis, the focusing efficiency of the metalens at different wavelengths was evaluated, as presented in Fig.~\ref{focusing_efficiency_allband}(a). Focusing efficiency is defined as the ratio of the effectively focused energy within the focal-plane region (calculated within a radius of $1.5\times\mathrm{FWHM}$ from the optical axis at the focal plane) to the total transmitted energy. 
 \edit{This definition corresponds to a transmission-normalized focusing efficiency, which characterizes the spatial energy confinement capability of the transmitted field and excludes reflection and absorption losses from the normalization. \editt{For reference, the corresponding incident-power-normalized focusing efficiency is obtained by multiplying this value by the total transmission (transmitted/incident power) at the same wavelength.}} The results show that the focusing efficiency remains moderate (20--38\%) across the 1800--2300~nm range and gradually increases with wavelength. This indicates that the device maintains reasonable energy confinement given the broad bandwidth and achromatic design requirements. In other words, achromatic focusing is achieved with an efficiency trade-off, which is commonly observed in broadband metalens systems. 
\editt{Imaging-relevant metrics are further assessed by the on-axis point-spread function (PSF), taken here as the simulated focal-plane intensity distribution at the design focus under plane-wave illumination. As a representative example, the on-axis PSF at $\boldsymbol{\lambda}=\mathbf{1.8}~\mu\mathrm{m}$ is shown in Fig.~\ref{focusing_efficiency_allband}(c), together with the corresponding polarization-vector map across the focal region. Further imaging characterization beyond the on-axis PSF reported here, such as field-dependent PSF/MTF across an image plane, off-axis aberrations, and full imaging demonstrations, is outside the scope of this study and will be addressed in future work.}

To examine overall spectral performance, Fig.~\ref{focusing_efficiency_allband}(b) also shows the total transmitted power through the metalens (regardless of focus) and the polarization purity (DoP) at the focal spot versus wavelength, highlighting the trade-off between these metrics. The transmission stays moderate level over 1.8--2.3~$\mu$m and increases slightly with wavelength, reaching $\sim$83\%. \edit{In this work, the DoP is defined based on the Stokes parameters as the ratio between the magnitude of the polarization vector and the total intensity, given by $\mathrm{DoP} = \sqrt{S_1^2 + S_2^2 + S_3^2} / S_0$, where the Stokes parameters are calculated from the simulated complex electric-field components at the focal plane.} 
We note that 
broadband focusing with reduced chromatic focal shift is the primary function of the proposed metasurface across 1.8--2.3~$\mu$m. We use geometric (PB) phase as an implementation route to realize the required spatial phase profile; polarization at the focus is not imposed as a design constraint here. Accordingly, polarization-related quantities (e.g., DoP) are reported here as auxiliary diagnostics rather than optimization objectives. 

Overall, the structure achieves 
moderate-to-high transmission (up to $\sim$83\%) indicates low loss; thus, the reduced focusing/polarization performance at the band edges is mainly due to phase-control limitations rather than absorption, and is likely associated with a structured (spatially varying) polarization state at focus (i.e., a vectorial focal field). 
These results provide a thorough characterization of the transmission--polarization trade-off in broadband metalens design. Future work could explore multi-layer metasurfaces or alternative geometries to better maintain polarization control across broad bandwidths and mitigate this trade-off.

\begin{figure}[h]
  \centering
  \includegraphics[width=0.95\textwidth]{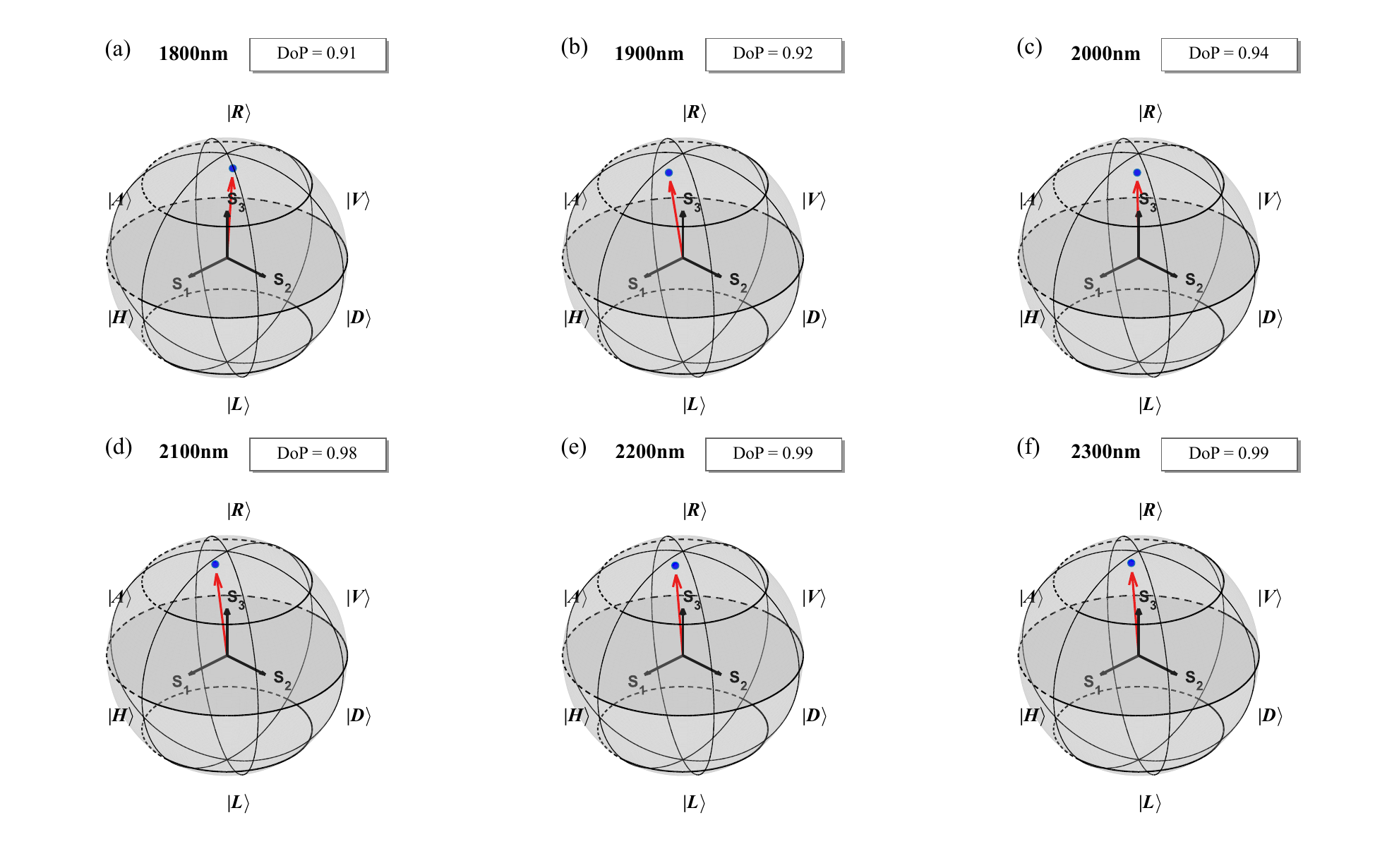}
  \caption{Simulated polarization states at the focal distance (over the focal region) visualized on the Poincar\'e sphere for wavelengths from 1800~nm to 2300~nm. Each red arrow represents the normalized Stokes vector of the focal-field polarization state, and the corresponding degree of polarization (DoP) is labeled beside each sphere.
  }
  \label{poincare_sphere}
  \end{figure}

For clarity in evaluating polarization control versus wavelength, Fig.~\ref{poincare_sphere} shows the output polarization states on the Poincar\'e sphere together with the corresponding DoP. The red vectors indicate the output Stokes-vector directions, and the blue dots denote their endpoints. \editt{The DoP quantifies polarization purity. As shown, the DoP remains within 0.94--0.99 over 2000-2300~nm, indicating relatively concentrated polarization states. As the wavelength approaches 1800~nm, the DoP drops slightly to a minimum of 0.91. This modest wavelength dependence is attributed to the vectorial nature of high-NA focusing, where the polarization varies across the focal region, together with the wavelength-dependent retardance of the anisotropic nanostructures; accordingly, the DoP here should be interpreted as a focal-field polarization-purity metric rather than a direct measure of conversion efficiency.}
\edit{The polarization response of anisotropic nanostructures based on the PB phase mechanism is inevitably wavelength-dependent. The phase delay between orthogonal polarization components may deviate from the ideal $\pi$ value at longer wavelengths, leading to incomplete polarization conversion. However, under high numerical aperture (NA) focusing, the optical field at the focal region can no longer be described by a single dominant polarization state. Instead, multiple polarization components coexist with pronounced spatial variation, giving rise to a more complex vectorial focal field. It should be emphasized that the reduction in DoP does not imply depolarization or increased optical loss. When calculating Stokes parameters across the entire focal region, such spatial polarization mixing manifests as a reduced DoP. The polarization distribution inset further visualizes the spatially varying polarization structure within the focal region, providing direct support for the above interpretation. This suggests that the polarization response has a limited operational bandwidth, although the detailed mechanism may require further investigation.}

While this study is based on numerical simulations, the proposed metalens design was developed with current nanofabrication capabilities in mind. The minimum feature width ($\sim$130~nm) is within the reach of electron-beam lithography or deep-UV lithography~\cite{Manfrinato2014}. The high aspect ratio (up to $\sim$13:1 for a 2~$\mu$m height) is challenging, but similar structures have been demonstrated~\cite{Duan2020, Pan2023}. The Si-on-CaF$_2$ platform provides low-loss SWIR performance and reasonable process compatibility, although additional support (e.g., carrier wafers) may be required during fabrication~\cite{Song2024}. Although dedicated tolerance simulations were not performed here, prior work indicates that metasurface devices, including achromatic metalenses, can remain robust under moderate fabrication deviations~\cite{Wang2022, Qiu2024}, supporting the practical feasibility of the proposed design.

\section*{\blue{Conclusion}}
This study presents, to the best of our knowledge, the first proposal and numerical validation of a broadband achromatic metalens operating in the 1800-2300~nm short-wave infrared (SWIR) band, constructed from a Si nanopillar array on a CaF$_2$ substrate.  
By combining geometric tuning with Pancharatnam–Berry (PB) phase engineering, continuous $2\pi$ phase modulation is achieved. The designed structure exhibits a stable focal \blue{ability}, \blue{broad operating range} capability, and low lateral aberrations across the entire operating band. The polarization analysis further demonstrates that as the wavelength increases, the degree of polarization (DoP) at the focal point shows a gradual upward trend. 
\edit{In addition, a wavelength-dependent evolution of the DoP analysis at the focus reveals the response of nanostructure anisotropy and vectorial nature of broadband focusing in high-numerical-aperture metalens systems , which lead to spatially nonuniform polarization-state superposition. These observations provide physical insight into polarization mixing effects at the focal spot.}
\edit{Overall, the results support the design of metalenses that achieve broadband achromatic focusing, while offering a framework for understanding polarization evolution in high-NA, broadband metasurface systems.} \blue{This work provides valuable design guidelines and a foundation for advanced photonic systems in the SWIR band that require spectral manipulation with reasonable transmission and focusing efficiencies. 
Future optimizations could build on this work to further improve polarization purity while maintaining broadband performance.} \blue{More work is needed to experimentally validate the proposed design, including fabrication and characterization of its broadband focusing and polarization performance.}

\vspace{6pt}
\noindent {\bf Funding:} This work was supported by UK Research and Innovation (UKRI) through the Engineering and Physical Sciences Research Council (EPSRC) Doctoral Training Partnership at the University of Glasgow [grant ref: EP/W524359/1]. This work was supported in part by the Royal Society (Grant No. RG\textbackslash R1\textbackslash 251474).

\noindent {\bf Contributions:} Y.H. performed the simulations and analysis, prepared the figures, and drafted the manuscript. A.C.D. provided supervision, contributed to methodology and data interpretation, and co‑wrote and revised the manuscript.

\noindent {\bf Acknowledgements:} We thank Natale Pruiti and Marc Sorel for their support with Lumerical simulations, and Shuhao Wu, Miles Padgett for helpful discussions.

\noindent {\bf Disclosures.}  The authors declare no conflicts of interest. 

\noindent {\bf Data availability.}  The datasets and code generated and/or analysed during the current study are available in the University of Glasgow Research Data Repository (DOI: 10.5525/gla.researchdata.2225), https://doi.org/10.5525/gla.researchdata.2225. 

\bibliography{references4}

\begin{thebibliography}{10}
\urlstyle{rm}
\expandafter\ifx\csname url\endcsname\relax
  \def\url#1{\texttt{#1}}\fi
\expandafter\ifx\csname urlprefix\endcsname\relax\def\urlprefix{URL }\fi
\expandafter\ifx\csname doiprefix\endcsname\relax\def\doiprefix{DOI: }\fi
\providecommand{\bibinfo}[2]{#2}
\providecommand{\eprint}[2][]{\url{#2}}

\bibitem{Mars2010}
\bibinfo{author}{Mars, J.~C.} \& \bibinfo{author}{Rowan, L.~C.}
\newblock \bibinfo{journal}{\bibinfo{title}{Spectral assessment of new aster
  swir surface reflectance data products for spectroscopic mapping of rocks and
  minerals}}.
\newblock {\emph{\JournalTitle{Remote Sensing of Environment}}}
  \textbf{\bibinfo{volume}{114}}, \bibinfo{pages}{2011--2025},
  \doiprefix\url{https://doi.org/10.1016/j.rse.2010.04.008}
  (\bibinfo{year}{2010}).

\bibitem{Campargue2017}
\bibinfo{author}{Campargue, A.} \emph{et~al.}
\newblock \bibinfo{journal}{\bibinfo{title}{The absorption spectrum of water
  vapor in the 2.2~$\mu$m transparency window: High sensitivity measurements
  and spectroscopic database}}.
\newblock {\emph{\JournalTitle{Journal of Quantitative Spectroscopy and
  Radiative Transfer}}} \textbf{\bibinfo{volume}{189}},
  \bibinfo{pages}{407--416},
  \doiprefix\url{https://doi.org/10.1016/j.jqsrt.2016.12.016}
  (\bibinfo{year}{2017}).

\bibitem{Knaeps2015}
\bibinfo{author}{Knaeps, E.} \emph{et~al.}
\newblock \bibinfo{journal}{\bibinfo{title}{A swir based algorithm to retrieve
  total suspended matter in extremely turbid waters}}.
\newblock {\emph{\JournalTitle{Remote Sensing of Environment}}}
  \textbf{\bibinfo{volume}{168}}, \bibinfo{pages}{66--79},
  \doiprefix\url{https://doi.org/10.1016/j.rse.2015.06.022}
  (\bibinfo{year}{2015}).

\bibitem{Pugliese2023}
\bibinfo{author}{Pugliese, E.}, \bibinfo{author}{Locatelli, M.},
  \bibinfo{author}{Meucci, R.}, \bibinfo{author}{Euzzor, S.} \&
  \bibinfo{author}{Poggi, P.}
\newblock \bibinfo{journal}{\bibinfo{title}{Swir digital holography and imaging
  through smoke and flames: unveiling the invisible}}.
\newblock {\emph{\JournalTitle{Optics Express}}} \textbf{\bibinfo{volume}{31}},
  \bibinfo{pages}{42090--42098},
  \doiprefix\url{https://doi.org/10.1364/oe.501602} (\bibinfo{year}{2023}).

\bibitem{Wen2018}
\bibinfo{author}{Wen, M.} \emph{et~al.}
\newblock \bibinfo{journal}{\bibinfo{title}{High-sensitivity short-wave
  infrared technology for thermal imaging}}.
\newblock {\emph{\JournalTitle{Infrared Physics \& Technology}}}
  \textbf{\bibinfo{volume}{95}}, \bibinfo{pages}{93--99},
  \doiprefix\url{https://doi.org/10.1016/j.infrared.2018.10.020}
  (\bibinfo{year}{2018}).

\bibitem{Attiaoui2020}
\bibinfo{author}{Attiaoui, A.} \emph{et~al.}
\newblock \bibinfo{journal}{\bibinfo{title}{Extended short-wave infrared
  absorption in group iv nanowire arrays}}.
\newblock {\emph{\JournalTitle{Physical Review Applied}}}
  \textbf{\bibinfo{volume}{15}},
  \doiprefix\url{https://doi.org/10.1103/physrevapplied.15.014034}
  (\bibinfo{year}{2020}).

\bibitem{Sarkar2020}
\bibinfo{author}{Sarkar, S.} \emph{et~al.}
\newblock \bibinfo{journal}{\bibinfo{title}{Short-wave infrared quantum dots
  with compact sizes as molecular probes for fluorescence microscopy}}.
\newblock {\emph{\JournalTitle{Journal of the American Chemical Society}}}
  \textbf{\bibinfo{volume}{142}}, \bibinfo{pages}{3449},
  \doiprefix\url{https://doi.org/10.1021/jacs.9b11567} (\bibinfo{year}{2020}).

\bibitem{Leemans2022}
\bibinfo{author}{Leemans, J.} \emph{et~al.}
\newblock \bibinfo{journal}{\bibinfo{title}{Colloidal iii–v quantum dot
  photodiodes for short-wave infrared photodetection}}.
\newblock {\emph{\JournalTitle{Advanced Science}}}
  \textbf{\bibinfo{volume}{9}}, \bibinfo{pages}{2200844},
  \doiprefix\url{https://doi.org/10.1002/advs.202200844}
  (\bibinfo{year}{2022}).

\bibitem{Wilson2015}
\bibinfo{author}{Wilson, R.~H.}, \bibinfo{author}{Nadeau, K.~P.},
  \bibinfo{author}{Jaworski, F.~B.}, \bibinfo{author}{Tromberg, B.~J.} \&
  \bibinfo{author}{Durkin, A.~J.}
\newblock \bibinfo{journal}{\bibinfo{title}{Review of short-wave infrared
  spectroscopy and imaging methods for biological tissue characterization}}.
\newblock {\emph{\JournalTitle{Journal of Biomedical Optics}}}
  \textbf{\bibinfo{volume}{20}}, \bibinfo{pages}{030901},
  \doiprefix\url{https://doi.org/10.1117/1.jbo.20.3.030901}
  (\bibinfo{year}{2015}).

\bibitem{Gutirrez-Gutirrez2025}
\bibinfo{author}{Gutiérrez-Gutiérrez, J.~A.}, \bibinfo{author}{Mieites, V.},
  \bibinfo{author}{López-Higuera, J.~M.} \& \bibinfo{author}{Conde, O.~M.}
\newblock \bibinfo{journal}{\bibinfo{title}{Rotating mirror short-wave infrared
  hyperspectral imaging system: Characterization and applications}}.
\newblock {\emph{\JournalTitle{Sensors and Actuators B: Chemical}}}
  \textbf{\bibinfo{volume}{439}}, \bibinfo{pages}{137762},
  \doiprefix\url{https://doi.org/10.1016/j.snb.2025.137762}
  (\bibinfo{year}{2025}).

\bibitem{Mehrubeoglu2020}
\bibinfo{author}{Mehrubeoglu, M.}, \bibinfo{author}{Sickle, A.~V.} \&
  \bibinfo{author}{Turner, J.}
\newblock \bibinfo{title}{Detection and identification of plastics using swir
  hyperspectral imaging}.
\newblock \bibinfo{pages}{15},
  \doiprefix\url{https://doi.org/10.1117/12.2570040}
  (\bibinfo{publisher}{SPIE-Intl Soc Optical Eng}, \bibinfo{year}{2020}).

\bibitem{Yao2025}
\bibinfo{author}{Yao, C.} \emph{et~al.}
\newblock \bibinfo{journal}{\bibinfo{title}{A stable and biocompatible
  shortwave infrared nanoribbon for dual-channel in vivo imaging}}.
\newblock {\emph{\JournalTitle{Nature Communications}}}
  \textbf{\bibinfo{volume}{16}}, \bibinfo{pages}{1--12},
  \doiprefix\url{https://doi.org/10.1038/s41467-024-55445-x}
  (\bibinfo{year}{2025}).

\bibitem{Abuzar2024}
\bibinfo{author}{Abuzar, M.}, \bibinfo{author}{Sheffield, K.} \&
  \bibinfo{author}{McAllister, A.}
\newblock \bibinfo{journal}{\bibinfo{title}{Feasibility of using
  swir-transformed reflectance (str) in place of surface temperature (ts) for
  the mapping of irrigated landcover}}.
\newblock {\emph{\JournalTitle{Land 2024, Vol. 13, Page 633}}}
  \textbf{\bibinfo{volume}{13}}, \bibinfo{pages}{633},
  \doiprefix\url{https://doi.org/10.3390/land13050633} (\bibinfo{year}{2024}).

\bibitem{Ustin2024}
\bibinfo{author}{Ustin, S.~L.} \& \bibinfo{author}{Middleton, E. M.~P.}
\newblock \bibinfo{journal}{\bibinfo{title}{Current and near-term
  earth-observing environmental satellites, their missions, characteristics,
  instruments, and applications}}.
\newblock {\emph{\JournalTitle{Sensors (Basel, Switzerland)}}}
  \textbf{\bibinfo{volume}{24}}, \bibinfo{pages}{3488},
  \doiprefix\url{https://doi.org/10.3390/s24113488} (\bibinfo{year}{2024}).

\bibitem{Jacob2022}
\bibinfo{author}{Jacob, D.~J.} \emph{et~al.}
\newblock \bibinfo{journal}{\bibinfo{title}{Quantifying methane emissions from
  the global scale down to point sources using satellite observations of
  atmospheric methane}}.
\newblock {\emph{\JournalTitle{Atmospheric Chemistry and Physics}}}
  \textbf{\bibinfo{volume}{22}}, \bibinfo{pages}{9617--9646},
  \doiprefix\url{https://doi.org/10.5194/acp-22-9617-2022}
  (\bibinfo{year}{2022}).

\bibitem{Gach2020}
\bibinfo{author}{Gach, J.~L.} \emph{et~al.}
\newblock \bibinfo{title}{C-red 3: A swir camera for fso applications}
  (\bibinfo{year}{2020}).

\bibitem{Prabhakar2020}
\bibinfo{author}{Prabhakar, S.} \emph{et~al.}
\newblock \bibinfo{journal}{\bibinfo{title}{Two-photon quantum interference and
  entanglement at 2.1 $\mu$m}}.
\newblock {\emph{\JournalTitle{Science Advances}}}
  \textbf{\bibinfo{volume}{6}}, \bibinfo{pages}{5195--5222},
  \doiprefix\url{https://doi.org/10.1126/sciadv.aay5195}
  (\bibinfo{year}{2020}).

\bibitem{Pirandola2020}
\bibinfo{author}{Pirandola, S.} \emph{et~al.}
\newblock \bibinfo{journal}{\bibinfo{title}{Advances in quantum cryptography}}.
\newblock {\emph{\JournalTitle{Advances in Optics and Photonics, Vol. 12, Issue
  4, pp. 1012-1236}}} \textbf{\bibinfo{volume}{12}},
  \bibinfo{pages}{1012--1236},
  \doiprefix\url{https://doi.org/10.1364/aop.361502} (\bibinfo{year}{2020}).

\bibitem{Bianconi2020}
\bibinfo{author}{Bianconi, S.} \& \bibinfo{author}{Mohseni, H.}
\newblock \bibinfo{journal}{\bibinfo{title}{Recent advances in infrared
  imagers: Toward thermodynamic and quantum limits of photon sensitivity}}.
\newblock {\emph{\JournalTitle{Reports on progress in physics. Physical Society
  (Great Britain)}}} \textbf{\bibinfo{volume}{83}}, \bibinfo{pages}{044101},
  \doiprefix\url{https://doi.org/10.1088/1361-6633/ab72e5}
  (\bibinfo{year}{2020}).

\bibitem{Khorasaninejad2017}
\bibinfo{author}{Khorasaninejad, M.} \& \bibinfo{author}{Capasso, F.}
\newblock \bibinfo{journal}{\bibinfo{title}{Metalenses: Versatile
  multifunctional photonic components}}.
\newblock {\emph{\JournalTitle{Science}}} \textbf{\bibinfo{volume}{358}},
  \doiprefix\url{https://doi.org/10.1126/science.aam8100}
  (\bibinfo{year}{2017}).

\bibitem{Arbabi2015}
\bibinfo{author}{Arbabi, A.}, \bibinfo{author}{Horie, Y.},
  \bibinfo{author}{Bagheri, M.} \& \bibinfo{author}{Faraon, A.}
\newblock \bibinfo{journal}{\bibinfo{title}{Dielectric metasurfaces for
  complete control of phase and polarization with subwavelength spatial
  resolution and high transmission}}.
\newblock {\emph{\JournalTitle{Nature Nanotechnology}}}
  \textbf{\bibinfo{volume}{10}}, \bibinfo{pages}{937--943},
  \doiprefix\url{https://doi.org/10.1038/nnano.2015.186}
  (\bibinfo{year}{2015}).

\bibitem{Wang2021}
\bibinfo{author}{Wang, Y.} \emph{et~al.}
\newblock \bibinfo{journal}{\bibinfo{title}{High-efficiency broadband
  achromatic metalens for near-ir biological imaging window}}.
\newblock {\emph{\JournalTitle{Nature Communications}}}
  \textbf{\bibinfo{volume}{12}}, \bibinfo{pages}{1--7},
  \doiprefix\url{https://doi.org/10.1038/s41467-021-25797-9}
  (\bibinfo{year}{2021}).

\bibitem{Chen2018}
\bibinfo{author}{Chen, W.~T.} \emph{et~al.}
\newblock \bibinfo{journal}{\bibinfo{title}{A broadband achromatic metalens for
  focusing and imaging in the visible}}.
\newblock {\emph{\JournalTitle{Nature Nanotechnology}}}
  \textbf{\bibinfo{volume}{13}}, \bibinfo{pages}{220--226},
  \doiprefix\url{https://doi.org/10.1038/s41565-017-0034-6}
  (\bibinfo{year}{2018}).

\bibitem{Liang2018}
\bibinfo{author}{Liang, H.} \emph{et~al.}
\newblock \bibinfo{journal}{\bibinfo{title}{Ultrahigh numerical aperture
  metalens at visible wavelengths}}.
\newblock {\emph{\JournalTitle{Nano Letters}}} \textbf{\bibinfo{volume}{18}},
  \bibinfo{pages}{4460--4466},
  \doiprefix\url{https://doi.org/10.1021/acs.nanolett.8b01570}
  (\bibinfo{year}{2018}).

\bibitem{Huang2021}
\bibinfo{author}{Huang, L.} \emph{et~al.}
\newblock \bibinfo{journal}{\bibinfo{title}{Long wavelength infrared imaging
  under ambient thermal radiation via an all-silicon metalens}}.
\newblock {\emph{\JournalTitle{Optical Materials Express, Vol. 11, Issue 9, pp.
  2907-2914}}} \textbf{\bibinfo{volume}{11}}, \bibinfo{pages}{2907--2914},
  \doiprefix\url{https://doi.org/10.1364/ome.434362} (\bibinfo{year}{2021}).

\bibitem{Chen2017}
\bibinfo{author}{Chen, B.~H.} \emph{et~al.}
\newblock \bibinfo{journal}{\bibinfo{title}{Gan metalens for pixel-level
  full-color routing at visible light}}.
\newblock {\emph{\JournalTitle{Nano Letters}}} \textbf{\bibinfo{volume}{17}},
  \bibinfo{pages}{6345--6352},
  \doiprefix\url{https://doi.org/10.1021/acs.nanolett.7b03135}
  (\bibinfo{year}{2017}).

\bibitem{Chen2012}
\bibinfo{author}{Chen, X.} \emph{et~al.}
\newblock \bibinfo{journal}{\bibinfo{title}{Dual-polarity plasmonic metalens
  for visible light}}.
\newblock {\emph{\JournalTitle{Nature Communications}}}
  \textbf{\bibinfo{volume}{3}}, \bibinfo{pages}{1--6},
  \doiprefix\url{https://doi.org/10.1038/ncomms2207} (\bibinfo{year}{2012}).

\bibitem{Wang2015}
\bibinfo{author}{Wang, W.} \emph{et~al.}
\newblock \bibinfo{journal}{\bibinfo{title}{Plasmonics metalens independent
  from the incident polarizations}}.
\newblock {\emph{\JournalTitle{Optics Express, Vol. 23, Issue 13, pp.
  16782-16791}}} \textbf{\bibinfo{volume}{23}}, \bibinfo{pages}{16782--16791},
  \doiprefix\url{https://doi.org/10.1364/oe.23.016782} (\bibinfo{year}{2015}).

\bibitem{Yu2019}
\bibinfo{author}{Yu, B.}, \bibinfo{author}{Wen, J.}, \bibinfo{author}{Chen, X.}
  \& \bibinfo{author}{Zhang, D.}
\newblock \bibinfo{journal}{\bibinfo{title}{An achromatic metalens in the
  near-infrared region with an array based on a single nano-rod unit}}.
\newblock {\emph{\JournalTitle{Applied Physics Express}}}
  \textbf{\bibinfo{volume}{12}}, \bibinfo{pages}{092003},
  \doiprefix\url{https://doi.org/10.7567/1882-0786/ab34c4}
  (\bibinfo{year}{2019}).

\bibitem{Zhang2020}
\bibinfo{author}{Zhang, Y.}, \bibinfo{author}{Yang, B.}, \bibinfo{author}{Liu,
  Z.} \& \bibinfo{author}{Fu, Y.}
\newblock \bibinfo{journal}{\bibinfo{title}{Polarization controlled dual
  functional reflective planar metalens in near infrared regime}}.
\newblock {\emph{\JournalTitle{Coatings 2020, Vol. 10, Page 389}}}
  \textbf{\bibinfo{volume}{10}}, \bibinfo{pages}{389},
  \doiprefix\url{https://doi.org/10.3390/coatings10040389}
  (\bibinfo{year}{2020}).

\bibitem{Xiao2023}
\bibinfo{author}{Xiao, S.} \emph{et~al.}
\newblock \bibinfo{journal}{\bibinfo{title}{Inverse design of a near-infrared
  metalens with an extended depth of focus based on double-process genetic
  algorithm optimization}}.
\newblock {\emph{\JournalTitle{Optics Express, Vol. 31, Issue 5, pp.
  8668-8681}}} \textbf{\bibinfo{volume}{31}}, \bibinfo{pages}{8668--8681},
  \doiprefix\url{https://doi.org/10.1364/oe.484471} (\bibinfo{year}{2023}).

\bibitem{Zou2025}
\bibinfo{author}{Zou, Y.}, \bibinfo{author}{Xu, Y.} \& \bibinfo{author}{Fang,
  M.}
\newblock \bibinfo{journal}{\bibinfo{title}{Broadband polarisation-insensitive
  metalens design with easy functionality expansion on an ingaas-detector
  substrate off-chip}}.
\newblock {\emph{\JournalTitle{SSRN Preprint}}} \bibinfo{pages}{ssrn.5221467},
  \doiprefix\url{https://doi.org/10.2139/ssrn.5221467} (\bibinfo{year}{2025}).

\bibitem{Zhang2016}
\bibinfo{author}{Zhang, S.} \emph{et~al.}
\newblock \bibinfo{journal}{\bibinfo{title}{High efficiency near
  diffraction-limited mid-infrared flat lenses based on metasurface
  reflectarrays}}.
\newblock {\emph{\JournalTitle{Optics Express, Vol. 24, Issue 16, pp.
  18024-18034}}} \textbf{\bibinfo{volume}{24}}, \bibinfo{pages}{18024--18034},
  \doiprefix\url{https://doi.org/10.1364/oe.24.018024} (\bibinfo{year}{2016}).

\bibitem{Guo2017}
\bibinfo{author}{Guo, Z.}, \bibinfo{author}{Tian, L.}, \bibinfo{author}{Shen,
  F.}, \bibinfo{author}{Zhou, H.} \& \bibinfo{author}{Guo, K.}
\newblock \bibinfo{journal}{\bibinfo{title}{Mid-infrared polarization devices
  based on the double-phase modulating dielectric metasurface}}.
\newblock {\emph{\JournalTitle{Journal of Physics D: Applied Physics}}}
  \textbf{\bibinfo{volume}{50}}, \bibinfo{pages}{254001},
  \doiprefix\url{https://doi.org/10.1088/1361-6463/aa6f9b}
  (\bibinfo{year}{2017}).

\bibitem{Ou2020}
\bibinfo{author}{Ou, K.} \emph{et~al.}
\newblock \bibinfo{journal}{\bibinfo{title}{Mid-infrared
  polarization-controlled broadband achromatic metadevice}}.
\newblock {\emph{\JournalTitle{Science Advances}}}
  \textbf{\bibinfo{volume}{6}}, \bibinfo{pages}{711--722},
  \doiprefix\url{https://doi.org/10.1126/sciadv.abc0711}
  (\bibinfo{year}{2020}).

\bibitem{Yue2023}
\bibinfo{author}{Yue, S.} \emph{et~al.}
\newblock \bibinfo{journal}{\bibinfo{title}{All-silicon
  polarization-independent broadband achromatic metalens designed for the
  mid-wave and long-wave infrared}}.
\newblock {\emph{\JournalTitle{Optics Express, Vol. 31, Issue 26, pp.
  44340-44352}}} \textbf{\bibinfo{volume}{31}}, \bibinfo{pages}{44340--44352},
  \doiprefix\url{https://doi.org/10.1364/oe.506471} (\bibinfo{year}{2023}).

\bibitem{Xu2025}
\bibinfo{author}{Xu, S.} \emph{et~al.}
\newblock \bibinfo{journal}{\bibinfo{title}{Inverse design broadband achromatic
  and polarizations-insensitive metalens in mid-infrared}}.
\newblock {\emph{\JournalTitle{Optics Communications}}}
  \textbf{\bibinfo{volume}{579}}, \bibinfo{pages}{131586},
  \doiprefix\url{https://doi.org/10.1016/j.optcom.2025.131586}
  (\bibinfo{year}{2025}).

\bibitem{Liu2024}
\bibinfo{author}{Liu, K.}, \bibinfo{author}{Sun, C.} \& \bibinfo{author}{Chui,
  H.~C.}
\newblock \bibinfo{journal}{\bibinfo{title}{Telecom-band high contrast
  narrowband metalens for 3d imaging}}.
\newblock {\emph{\JournalTitle{Optics and Lasers in Engineering}}}
  \textbf{\bibinfo{volume}{180}}, \bibinfo{pages}{108325},
  \doiprefix\url{https://doi.org/10.1016/j.optlaseng.2024.108325}
  (\bibinfo{year}{2024}).

\bibitem{Zhang2025-1}
\bibinfo{author}{Zhang, F.} \emph{et~al.}
\newblock \bibinfo{journal}{\bibinfo{title}{Dispersion-engineered spin
  photonics based on folded-path metasurfaces}}.
\newblock {\emph{\JournalTitle{Light: Science and Applications}}}
  \textbf{\bibinfo{volume}{14}}, \bibinfo{pages}{1--10},
  \doiprefix\url{https://doi.org/10.1038/s41377-025-01850-w}
  (\bibinfo{year}{2025}).

\bibitem{Zhang2025-2}
\bibinfo{author}{Zhang, Y.} \emph{et~al.}
\newblock \bibinfo{journal}{\bibinfo{title}{On-chip integration of achromatic
  metalens arrays}}.
\newblock {\emph{\JournalTitle{Nature Communications}}}
  \textbf{\bibinfo{volume}{16}}, \bibinfo{pages}{1--8},
  \doiprefix\url{https://doi.org/10.1038/s41467-025-62539-7}
  (\bibinfo{year}{2025}).

\bibitem{Aieta2015}
\bibinfo{author}{Aieta, F.}, \bibinfo{author}{Kats, M.~A.},
  \bibinfo{author}{Genevet, P.} \& \bibinfo{author}{Capasso, F.}
\newblock \bibinfo{journal}{\bibinfo{title}{Multiwavelength achromatic
  metasurfaces by dispersive phase compensation}}.
\newblock {\emph{\JournalTitle{Science}}} \textbf{\bibinfo{volume}{347}},
  \bibinfo{pages}{1342--1345},
  \doiprefix\url{https://doi.org/10.1126/science.aaa2494}
  (\bibinfo{year}{2015}).

\bibitem{Wang2018}
\bibinfo{author}{Wang, S.} \emph{et~al.}
\newblock \bibinfo{journal}{\bibinfo{title}{A broadband achromatic metalens in
  the visible}}.
\newblock {\emph{\JournalTitle{Nature Nanotechnology}}}
  \textbf{\bibinfo{volume}{13}}, \bibinfo{pages}{227--232},
  \doiprefix\url{https://doi.org/10.1038/s41565-017-0052-4}
  (\bibinfo{year}{2018}).

\bibitem{Guo2022}
\bibinfo{author}{Guo, K.}, \bibinfo{author}{Wang, C.}, \bibinfo{author}{Kang,
  Q.}, \bibinfo{author}{Chen, L.} \& \bibinfo{author}{Guo, Z.}
\newblock \bibinfo{journal}{\bibinfo{title}{Broadband achromatic metalens with
  polarization insensitivity in the mid-infrared range}}.
\newblock {\emph{\JournalTitle{Optical Materials}}}
  \textbf{\bibinfo{volume}{131}}, \bibinfo{pages}{112489},
  \doiprefix\url{https://doi.org/10.1016/j.optmat.2022.112489}
  (\bibinfo{year}{2022}).

\bibitem{Chen2019}
\bibinfo{author}{Chen, W.~T.}, \bibinfo{author}{Zhu, A.~Y.},
  \bibinfo{author}{Sisler, J.}, \bibinfo{author}{Bharwani, Z.} \&
  \bibinfo{author}{Capasso, F.}
\newblock \bibinfo{journal}{\bibinfo{title}{A broadband achromatic
  polarization-insensitive metalens consisting of anisotropic nanostructures}}.
\newblock {\emph{\JournalTitle{Nature Communications}}}
  \textbf{\bibinfo{volume}{10}}, \bibinfo{pages}{1--7},
  \doiprefix\url{https://doi.org/10.1038/s41467-019-08305-y}
  (\bibinfo{year}{2019}).

\bibitem{Manfrinato2014}
\bibinfo{author}{Manfrinato, V.~R.} \emph{et~al.}
\newblock \bibinfo{journal}{\bibinfo{title}{Determining the resolution limits
  of electron-beam lithography: Direct measurement of the point-spread
  function}}.
\newblock {\emph{\JournalTitle{Nano Letters}}} \textbf{\bibinfo{volume}{14}},
  \bibinfo{pages}{4406--4412},
  \doiprefix\url{https://doi.org/10.1021/nl5013773} (\bibinfo{year}{2014}).

\bibitem{Duan2020}
\bibinfo{author}{Duan, T.}, \bibinfo{author}{Gu, C.}, \bibinfo{author}{Ang,
  D.~S.}, \bibinfo{author}{Xu, K.} \& \bibinfo{author}{Liu, Z.}
\newblock \bibinfo{journal}{\bibinfo{title}{A novel fabrication technique for
  high-aspect-ratio nanopillar arrays for sers application}}.
\newblock {\emph{\JournalTitle{RSC Advances}}} \textbf{\bibinfo{volume}{10}},
  \bibinfo{pages}{45037--45041},
  \doiprefix\url{https://doi.org/10.1039/d0ra09145f} (\bibinfo{year}{2020}).

\bibitem{Pan2023}
\bibinfo{author}{Pan, A.} \emph{et~al.}
\newblock \bibinfo{journal}{\bibinfo{title}{Fabrication of ultrahigh aspect
  ratio si nanopillar and nanocone arrays}}.
\newblock {\emph{\JournalTitle{Journal of Vacuum Science \& Technology B}}}
  \textbf{\bibinfo{volume}{41}}, \bibinfo{pages}{23001},
  \doiprefix\url{https://doi.org/10.1116/6.0002276} (\bibinfo{year}{2023}).

\bibitem{Song2024}
\bibinfo{author}{Song, Q.} \emph{et~al.}
\newblock \bibinfo{journal}{\bibinfo{title}{Wafer-scale fabrication of
  single-crystalline calcium fluoride thin-film on insulator by ion-cutting}}.
\newblock {\emph{\JournalTitle{Optical Materials}}}
  \textbf{\bibinfo{volume}{157}}, \bibinfo{pages}{115787},
  \doiprefix\url{https://doi.org/10.1016/j.optmat.2024.115787}
  (\bibinfo{year}{2024}).

\bibitem{Wang2022}
\bibinfo{author}{Wang, H.~C.}, \bibinfo{author}{Achouri, K.} \&
  \bibinfo{author}{Martin, O.~J.}
\newblock \bibinfo{journal}{\bibinfo{title}{Robustness analysis of
  metasurfaces: Perfect structures are not always the best}}.
\newblock {\emph{\JournalTitle{ACS Photonics}}} \textbf{\bibinfo{volume}{9}},
  \bibinfo{pages}{2438--2447},
  \doiprefix\url{https://doi.org/10.1021/acsphotonics.2c00563}
  (\bibinfo{year}{2022}).

\bibitem{Qiu2024}
\bibinfo{author}{Qiu, Y.}, \bibinfo{author}{Deng, L.}, \bibinfo{author}{Zhan,
  Y.}, \bibinfo{author}{Li, G.} \& \bibinfo{author}{Guan, J.}
\newblock \bibinfo{journal}{\bibinfo{title}{The effect of height error on
  performance of propagation phase-based metalens}}.
\newblock {\emph{\JournalTitle{Micromachines 2024, Vol. 15, Page 540}}}
  \textbf{\bibinfo{volume}{15}}, \bibinfo{pages}{540},
  \doiprefix\url{https://doi.org/10.3390/mi15040540} (\bibinfo{year}{2024}).

\end{thebibliography}

\end{document}